\documentstyle[prb,aps,twocolumn,floats]{revtex}
\begin{document}
\draft
\title{Quantum ballistic transport in constrictions of n-PbTe}
\author{G. Grabecki, J. Wr\'obel, T. Dietl}
\address{Institute of Physics, Polish Academy of Sciences,
al. Lotnik\'ow 32/46, PL-02-668 Warszawa, Poland}
\author{K. Byczuk}
\address{Institute of Theoretical Physics,
Warsaw University, PL-00-681 Warszawa, Poland}
\author{E. Papis, E. Kami\'nska, A. Piotrowska}
\address{Institute of Electron Technology,
PL-02-668 Warszawa, Poland\\}
\author{G. Springholz, M. Pinczolits, and G. Bauer}
\address{Institut f\"ur Halbleiterphysik,
Johannes Kepler Universit\"at Linz, A-4040 Linz, Austria}

\maketitle

\begin{abstract}
Conductance of submicron constrictions of PbTe:Bi was studied up
to 8~T and between 4.2~K and 50~mK. The structures were fabricated
by electron beam lithography and chemical etching of
high--electron mobility films grown by MBE on BaF$_2$.  In the
moderately strong magnetic fields perpendicular to the current, $B
~{\ge}$ 1~T, the conductance shows accurate quantization in the
units of 1$e^2/h$ as a function of the side--gate voltage.  In the
absence of the field, a temperature--independent step structure,
with an average step height $\approx e^2/h$, is observed. It is
suggested that such a quantization may reflect the lifting of the
Kramers degeneracy by the exchange interaction among the
electrons, effective despite a large dielectric constant of bulk
PbTe.
\end{abstract}

\preprint{Dec. 20 '98}
\narrowtext

Recent studies on point contacts and wires of high mobility
GaAs/AlGaAs heterostructures suggest that conductance quantization
deviates from a simple behavior expected for electron ballistic
transport in one-dimensional systems. In particular, Thomas {\em
et al.}\cite{Thom96} have detected a conductance step at $\sim
1.4e^2/h$, which evolves in a strong in-plane magnetic field into
the spin-split plateau at $e^2/h$. The spin-orbit
coupling,\cite{Thom96} a spontaneous spin-splitting driven by the
exchange interaction within the electron
liquid,\cite{Thom96,Gold96,Wang96} or scattering by
plasmons\cite{Kris98} have been considered as possible mechanisms
accounting for the observation. Similar findings have been
reported for structures characterized by somewhat lower values of
electron mobilities.\cite{Tsch96,Ramv97} In longer wires,  as
shown by Tarucha {\em et al.}\cite{Taru95} and Yacoby {\em et
al.},\cite{Yaco96} the conductance steps assume the quantized
values $n2e^2/h$ at high temperatures, $T \sim 1$ K, but on
lowering temperature their magnitude gradually decreases, $G =
rn2e^2/h$, where $r < 1$. A combined influence of disorder,
Luttinger-liquid effects, and coupling to the Fermi-liquid
reservoirs is thought\cite{Yaco96,Masl95,Alek98} to result in such
a behavior of the conductance.

Owing to the high electron mobility $\mu$ at low temperatures,
n-PbTe constitutes a promising material for studies of 1D systems
by means of ballistic transport. The large magnitudes of $\mu$,
$10^5 - 10^6$ cm$^2$/Vs, achieved with no modulation doping,
result from the low values of the effective masses
($m^*_{\perp}/m_o \approx 0.02, m^*_{\parallel}/m_o \approx 0.2$),
and the high value of the static dielectric constant, $\epsilon_L
\approx 10^3$. The latter stems from the fact that PbTe is at the
border line to a ferroelectric phase transition due to a
cubic-rhombohedral distortion.\cite{Baue83} The large value of
$\epsilon_L$ reduces significantly the magnitude of backward
scattering by the Coulomb potentials. We suggest that this is the
main reason for novel aspects of ballistic transport in n-PbTe
reported in the present paper.  In particular, the destructive
effect of electron scattering upon quantization of the density of
states is much weaker, for a given value of $\mu$, in the case of
IV-VI compounds than for standard
semiconductors.\cite{vanW91,Tobb95,Ensslin98} Indeed, in the
latter, the lifetime $\tau_q$ of quantum states is much shorter
than the momentum relaxation time $\tau$ calculated from the
mobility.\cite{Diet78} By contrast, since in IV-VI compounds the
Coulomb potentials are screened, electron scattering is dominated
by short range part of defect potentials, for which $\tau_q
\approx \tau$.

We have grown by MBE $0.7$ $\mu$m thick epilayers of PbTe:Bi onto
\{111\}-oriented BaF$_2$ substrates. Because of the
difference between thermal expansion coefficients of BaF$_2$ and
PbTe, the epilayers are under tensile strain at low
temperatures. As a result, the fourfold valley degeneracy at
the L point of the Brillouin zone is lifted and
the valley with its main
axis oriented along $\langle111\rangle$ is shifted downwards in
energy as compared to the three obliquely oriented
valleys.\cite{Sing86,Grab97}
Furthermore, the interface defects, brought about by both mismatch
of the lattice constants and the thermal stresses, limit the
electron mobility and deplete the vicinity of the interface
from the carriers.\cite{Ueta97}  From the magnitudes of the Hall
resistance and electron concentration $n = 2\times10^{17}$
cm$^{-3}$ resulting from the period of the Shubnikov--de Haas
oscillations we evaluate the width of the depletion
region to be between 0.4 and 0.5 $\mu$m in our layers.
The value of $n$, together with the Hall mobility $\mu_H =
2\times10^{5}$ cm$^2$/Vs, lead to the mean free path $\ell =
1.6$ $\mu$m at 4.2~K.

Submicron constrictions with side gates have been fabricated of
the epilayers employing single--level electron--beam lithography
followed by wet chemical etching. This process produces a
semicircular constriction of geometrical radius of about 0.5 $\mu
$m in the most etched region, as shown in Fig.~1. This is
comparable to the width of the depletion layer implying that the
carriers occupy only a narrow channel located near the top of the
constriction. Further reduction of the channel width can be
achieved by means of the two side PbTe gates. Since the dielectric
constant of BaF$_2$ is about 8, the capacitive coupling between
the gates and the channel proceeds mainly {\it via} the substrate.
Accordingly, the electrons are pushed towards the top of the
constriction for the negative gate voltages.  We have noted that
the pinch--off voltage, $V_{th}$, slowly evolves with time, even
at low temperatures, as well as it becomes more negative after
subsequent cooling cycles. These effects are presumably caused by
slow and incomplete strain relaxation at the interface. We have
checked that there is no gate leakage current up to the gate
voltage $V_g$ = ${\pm}$200 V, which implies that the dielectric
strength of BaF$_2$ is $E_s \approx 150$ V/$\mu$m. For the
conductance measurements, low-frequency a.c. currents of
magnitudes not greater than 2 nA and a phase sensitive detection
have been employed.

Figure 2 presents the conductance $G=1/R$ as a function of gate
voltage in units of $e^2/h$ at various values of the magnetic
field ${\bf B}$ for its two directions in the plane perpendicular
to the current. Several conclusions emerge from those findings.
First, when depleting the electrons in the constriction by $V_g$,
a sequence of conductance steps is observed, whose width increases
strongly with $B$, an observation confirming the strong
suppression of backscattering and the depopulation of 1D subbands
by the magnetic field.\cite{Been91} Second, while the steps are
accurately quantized for the in-plane magnetic field, the plateau
values are systematically greater than multiples of $e^2/h$ in the
case of ${\bf B}$ along the growth axis. This is consistent with
the fact that for our four--probe arrangement shown in Fig.~1, the
influence of series resistances is negligible, whereas the Hall
effect of the wide regions gives a contribution to
$R$.\cite{Been91} Actually, as shown in Fig.~3, the deviations of
$R$ from the quantized values are linear in $B$, and corresponds
to a value of the apparent Hall constant, which is by a factor of
about two greater than that measured prior to the nanofabrication.
Finally, the rate of change of the filling factor, d$\nu$/d$V_g$
is about two times larger if ${\bf B}$ is in the plane of the
epilayer comparing to the case of ${\bf B}$ along the growth
direction. This means that the ratio of an effective width $w$ of
the constriction to its height $d$ is about 2.

As already mention, we presume that the gate voltage-induced change
of the charge in the construction is primary associated with a
change in the width of the depletion region. In order to evaluate
the constriction cross-section $S\approx 2d^2$ at given $V_g$,
we employ the modified Sharvin relation,\cite{Torr94} generalized
to the case of the ellipsoidal Fermi surface,
\begin{equation}
G(d) = \frac{e^2}{h} \frac{(3 \pi ^2n)^{2/3}}{2 \pi}
(\frac{m_{\parallel}^*}{m_{\perp}^*
})^{1/6}(2d^2)(1-\frac{1}{\sqrt{2}(3 \pi^2n)^{1/3}d})^2.
\end{equation}
In this way we estimate $d\approx 15$ nm for $n =2\times10^{17}$
cm$^{-3}$ and  $G = e^2/h$. This value of $d$, together with the
effective mass $m^* = 0.02m_o$, leads to the energy distance
between the ground-state and excited electric subbands of about 19
meV.\cite{Sche96} It is worth noting that despite a substantial
increase in the density of the electric subbands with $d$, and
thus with $V_g - V_{th}$, no corresponding degradation of
conductance quantization is observed. This follows presumably from
a reduction of scattering when the electrons becomes less squeezed
near the top surface.

Of particular relevance are experimental results in the absence
of the magnetic field, presented in an expanded scale in Fig.~4.
Distinct and temperature--independent
step-like structure, superimposed on a smoothly increasing
background, is visible.  The lack of temperature--dependent
resonances or fluctuations suggest that the constriction is both
adiabatic and non-diffusive.  However, many plateaus do not occur
at the quantized values, and the conductance difference between them
corresponds rather to $e^2/h$, not to $2e^2/h$.  While there are
clearly visible changes in the dependence of the conductance
on the gate voltage after subsequent cooling cycles, the above
conclusion appears to remain valid.

The most natural explanation of these results would be the
assumption that the structure corresponds to a {\em
quasi}-ballistic regime, $L \approx \ell$, in which impurity
scattering may reduce the width and height of the
steps.\cite{Hann89} At the same time we note that the spin-orbit
coupling, although rather strong in PbTe, is not expected to lift
the two-fold degeneracy according to the Kramers theorem. It might
appear also that the large dielectric constant of PbTe makes
effects of electron-electron interactions totally unimportant.

We have, however, decided to develop a Hartree-Fock theory of the
electron--electron interaction for a quantum wire with the
dielectric constant $\epsilon_1$ embedded in an environment with
the dielectric constant $\epsilon_2$.\cite{Bycz99}  The wire is
modeled by an infinitely long cylinder of the radius $a$.   An
electrostatic potential between two electrons is determined by
solving the Poisson equation with the appropriate boundary
conditions at the cylinder boundary.  We have obtained the matrix
element $V(q)$ of the interaction potential in the one-particle
basis of states corresponding to the ground state subband. While
for $qa \gg 1$ $V(q) \sim 1/\epsilon_1 q^2$, in the opposite limit
$V(q) = -2e^2\ln (aq/2)/\epsilon_2$; so it diverges
logarithmically.  This behavior, independent of the exact shape of
the subband wave function, is characteristic for a 1D Fourier
transform of the Coulomb interaction $e^2/r$ (with $a$ as a
short-distance cut off).\cite{Gold96}  Moreover, it turns out that
$V(q)$ for $ aq \rightarrow 0$ is entirely determined by the
interaction with the image charges as it does not depend on the
dielectric constant $\epsilon_1$ of the wire but only on
$\epsilon_2$.  This means that in the structures under
consideration, for which a substantial part of the image charge
resides in either vacuum or BaF$_2$, the exchange energy $E_{ex}$
is rather large since it is determined by $V(q)$ for $q
\rightarrow 0$.

Within the above model, we have computed a critical density $n_c$,
below which the Hartree-Fock ground state of the 1D electron
liquid is ferromagnetic.  By taking $\epsilon_1 = 1350$,
$\epsilon_2 = 1$, $m^* = 0.02m_o$, and $a =d/\sqrt{2} =8.5$ nm we
obtain $n_c = 0.64{\cdot}10^5$ cm$^{-1}$.  On the other hand, the
concentration of the spin-polarized electrons filling up the
ground-state subband till the bottom of the first excited subband
is $n_{1d} = \sqrt{2m^*E_1}/h = 1.6 \times10^5$ cm$^{-1}$, as $E_1
\approx 19 $ meV.  Those estimates suggest, therefore, that over a
significant range of the gate voltage, the concentration of the
electrons in particular subbands is small enough for the
spin-polarization to be sustained. The tendency toward a
ferromagnetic ordering can be enhanced even further in the case of
chaotic level distribution in low-dimensional
systems.\cite{Andr98}

In conclusion, we have fabricated
submicron constrictions of PbTe coupled to 3D reservoirs.
Magnetoconductance measurements show accurately
quantized steps for any direction of the magnetic field perpendicular to
the current.  At the same time, our findings suggest that the spin
degeneracy is removed in the absence of the magnetic field. While
the Hartree-Fock model is not expected to provide the accurate
value of the instability point, and the actual ground state can be
more complex than a simple ferromagnetic phase, our result
demonstrate that despite the large dielectric constant,
electron-electron interaction may account for the observed
behavior of the ballistic conductance.

Poland as well as FWF Vienna, ``\"{O}sterreichische
Nationalbank'', Ost-West Funds, and  GME in Austria for financial
support.

\begin{figure}
\caption[]{SEM picture of a constriction made of PbTe:Bi (white regions).
AFM depth profiles along (X) and across (Y) the constriction are
shown in right panels. The horizontal scales correspond exactly to
the magnification of the SEM picture. }
\end{figure}

\begin{figure}
\caption[]{Conductance as a function of the gate voltage in PbTe
constriction for two configurations of the magnetic field
perpendicular to the current a){\bf B}${\perp}\langle111\rangle$,
b){\bf B}${\parallel}\langle111\rangle$ Numbers denote subsequent
values of the magnetic field. Cross--sections of the constrictions
together with positions of the edge currents (shaded area) are
shown schematically.}
\end{figure}

\begin{figure}
\caption[]{Resistance of the PbTe constriction as a function
of the magnetic field at 50~mK. The difference
between the measured (solid lines) and quantized values
(dotted lines) is given by solid dots. The dashed line corresponds
to the resulting Hall resistance of the wide regions.}
\end{figure}

\begin{figure}
\caption[]{Conductance of a PbTe constriction at various
temperatures as a function of the gate voltage.}
\end{figure}
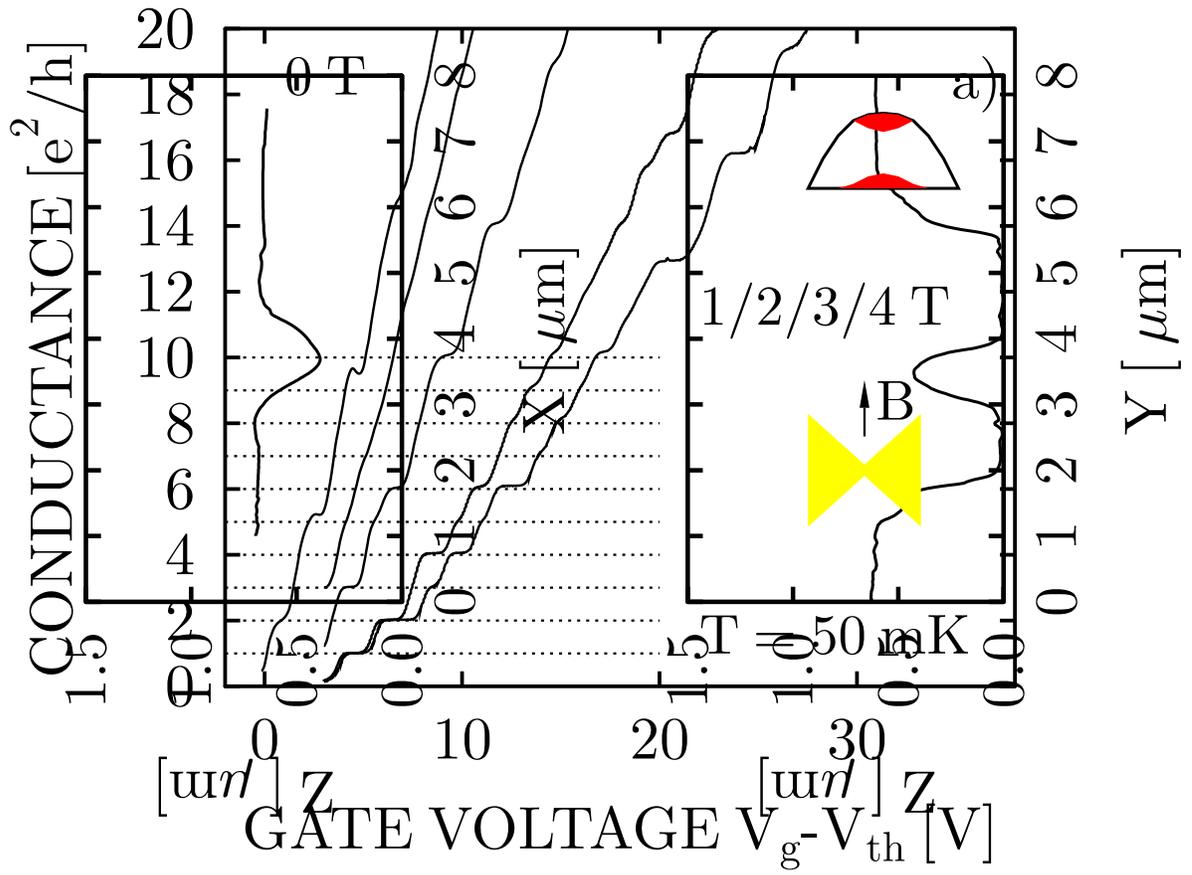
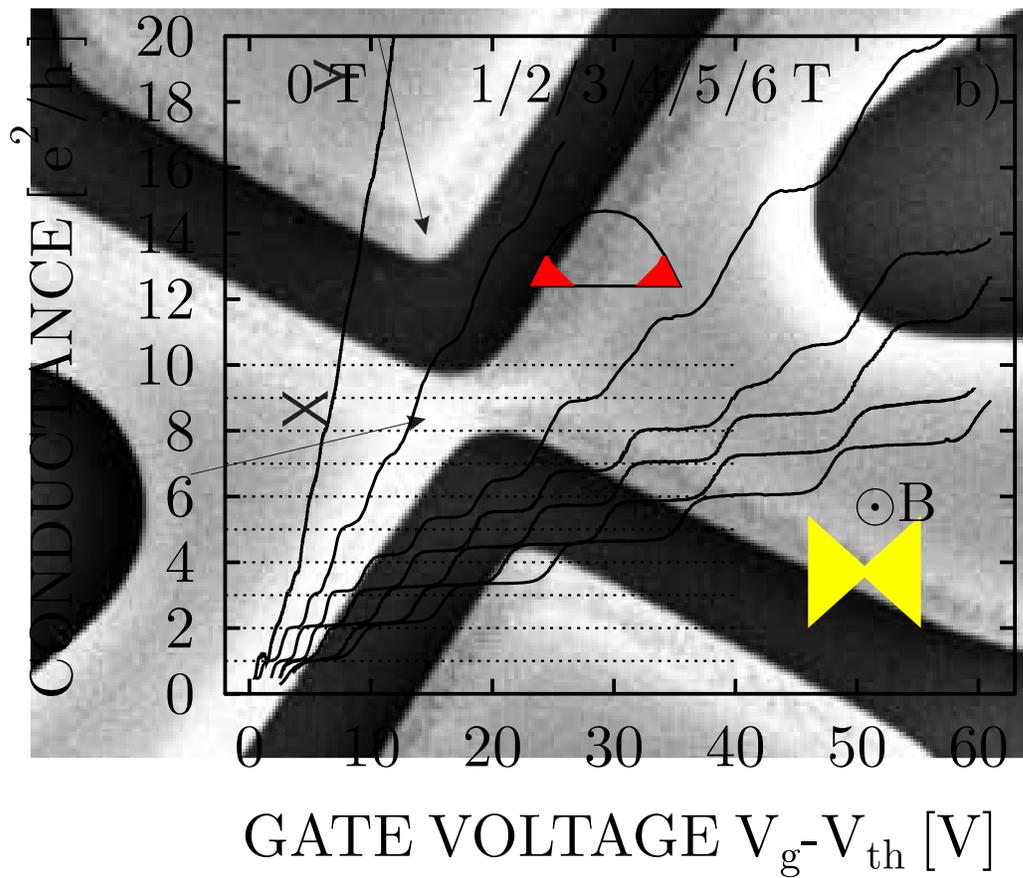
\end{document}